# Ensemble Learning Based Convex Approximation of Three-Phase Power Flow

Ren Hu, Qifeng Li, *Member, IEEE*, Feng Qiu, *Senior Member, IEEE*,

*Abstract*—Though the convex optimization has been widely used in power systems, it still cannot guarantee to yield a tight (accurate) solution to some problems. To mitigate this issue, this paper proposes an ensemble learning based convex approximation for AC power flow equations that differs from the existing convex relaxations. The proposed approach is based on quadratic power flow equations in rectangular coordinates and it can be used in both balanced and unbalanced three-phase power networks. To develop this data-driven convex approximation of power flows, the polynomial regression (PR) is first deployed as a basic learner to fit convex relationships between the independent and dependent variables. Then, ensemble learning algorithms such as gradient boosting (GB) and bagging are introduced to combine learners to boost model performance. Based on the learned convex approximation of power flows, optimal power flow (OPF) is formulated as a convex quadratic programming problem. The simulation results on IEEE standard cases show that, in the context of solving OPF, the proposed data-driven convex approximation outperforms the conventional SDP relaxation in both accuracy and computational efficiency, especially in the cases that the conventional SDP relaxation fails.

*Index Terms*-- convex approximation, data-driven, ensemble learning, power flow

## I. INTRODUCTION

POWER flow analysis plays a significant role in power system planning and operation. Many decision-making processes in power systems rely heavily on accurate and effective power flow calculations [1]-[3]. Power flow models also function as inevitable system constraints in optimization problems like transmission or generation expansion planning and optimal power flow (OPF) [3]-[5]. However, the nonlinearity and nonconvexity of AC power flow (PF) make the OPF problem computationally expensive due to the NP-hardness.

One approach to handling these challenges is to linearize the power flow equations, which has been widely adopted in power system dispatching [6] [7] and power market trading [8] [9]. One of the most well-known linear PF models, the DC power flow, captures the linear relationship between the active power flow injection and the bus voltage phase angle. Other extended versions of linear power flow, that involves reactive power, have also attracted substantial attention [10]-[14]. Although linear power flow models are computationally tractable, they are generally based on some critical assumptions, such as ignoring the inherent interactions between bus voltages. For the sake of a better predictive accuracy, some learning-based linear models utilize principal component analysis (PCA) based methods to narrow the bus voltage (independent) variable space by replacing the original independent variables by a new small subset of variables. This indirectly changes the bus number of the power system and introduces poor applicability and interpretability into further control and optimization applications.

With the deluge of data generated by sensors like phasor measurement units today, data-driven methods have attracted massive research efforts in power system analysis, such as in estimating distribution factors [15] and the Jacobian matrix [16] and identifying the admittance matrix [17]. In this paper, machine learning approaches will be applied to develop convex approximation of ACPF. Through data mining methods, we obtain the convex approximation of the original power flow that considers inherent interactive terms between bus voltages and maintains the bus number. More precisely, the polynomial regression [18] [19] is employed as a basic learner to infer convex relationships between the active or reactive power and the bus voltage, given that their functional formulations are originally nonconvex. Ensemble learning algorithms, i.e., gradient boosting [20] [21] and bagging [22]-[24], are then introduced to assemble every basic learner at each iteration, tune the regularization parameters to avoid overfitting, and eventually yield a stronger learner.

Although nonlinear programming problems are generally NP-hard to solve, many convex nonlinear optimization problems admit polynomial-time algorithms [25]. In recent years, high-performance solvers such as MOSEK, CPLEX, and GUROBI have been developed to effectively solve major types of convex problems. Various convex relaxations, such as second-order cone (SOC) [26], semi-definite programming (SDP) [27], enhanced SDP [28] convex distflow (CDF) [29], [30], quadratic convex (QC) [31], moment-based [32], and convex hull relaxation [33], have been introduced to convexify a fundamental power system optimization problem (OPF). Of these, SDP-based relaxations have attracted the most attention due to its general applicability to nonconvex quadratic problems [30]-[32] including the three-phase ACPF investigated in this paper. However, they are generally computationally hard and not tight enough to guarantee a satisfactory solution for many OPF cases and result in failures [36], [44], [45]. The SOC relaxation is computationally easier and, however, not as tight as SDP [33], while CDF relaxation is only suitable for balanced tree networks. To avoid the

.

limitations of the existing convex relaxations, this paper develops a novel convex approximation of power flow models through using data-driven methods. The main contributions of this paper can be summarized as follows:

- A data-driven convex quadratic approximation (DDCQA) of power flow is proposed and applied to convexify the OPF. The resulting OPF is a convex quadratic programming problem that outperforms the existing SDP relaxation in computational efficiency. More importantly, the accuracy of the proposed DDCQA can be improved by learning from operation experience.
- In the parameter-fitting process, ensemble learning algorithms are applied to incorporate all basic learners, i.e. the polynomial regression, into a stronger learner in order to enhance the predictive accuracy of DDCQA.

The rest of this paper is organized as follows: Section II depicts the existing problems and proposed solution for computing and fitting power flow. In Section III, data-driven convexification of power flow is formulated through ensemble learning. The empirical IEEE case analyses and conclusions are displayed in Section IV and V, respectively.

## II. PROBLEM FORMULATION AND STATEMENT

This section first presents a three-phase ACPF models in rectangular coordinates. Then, the existing problems are discussed and an overview of the proposed method is provided.

### A. Three-Phase AC Power Flows in Rectangular Coordinates

In an *n*-bus three-phase power networks, the equations of ACPF in rectangular coordinates for each phase involving the coupling terms from another two phases can be generally illustrated as below.

$$
\begin{aligned}
P_i^\phi &= e_i^\phi \sum_{j\in N} \sum_{\gamma\in\Phi} \left(G_{ij}^{\phi\gamma} e_j^\gamma - B_{ij}^{\phi\gamma} f_j^\gamma\right) + \\
&\quad f_i^\phi \sum_{j\in N} \sum_{\gamma\in\Phi} (G_{ij}^{\phi\gamma} f_j^\gamma + B_{ij}^{\phi\gamma} e_j^\gamma) \\
Q_i^\phi &= f_i^\phi \sum_{j\in N} \sum_{\gamma\in\Phi} \left(G_{ij}^{\phi\gamma} e_j^\gamma - B_{ij}^{\phi\gamma} f_j^\gamma\right) - \\
&\quad e_i^\phi \sum_{j\in N} \sum_{\gamma\in\Phi} (G_{ij}^{\phi\gamma} f_j^\gamma + B_{ij}^{\phi\gamma} e_j^\gamma) \\
P_{ij}^\phi &= e_i^\phi \sum_{\gamma\in\Phi} \left[G_{ij}^{\phi\gamma}(e_i^\gamma - e_j^\gamma) + B_{ij}^{\phi\gamma}\left(f_j^\gamma - f_i^\gamma\right)\right] + \\
&\quad f_i^\phi \sum_{\gamma\in\Phi} [G_{ij}^{\phi\gamma}(f_i^\gamma - f_j^\gamma) + B_{ij}^{\phi\gamma}\left(e_i^\gamma - e_j^\gamma\right)] \\
Q_{ij}^\phi &= e_i^\phi \sum_{\gamma\in\Phi} [G_{ij}^{\phi\gamma}(f_j^\gamma - f_i^\gamma) - B_{ij}^{\phi\gamma}\left(e_i^\gamma - e_j^\gamma\right)] + \\
&\quad f_i^\phi \sum_{\gamma\in\Phi} [G_{ij}^{\phi\gamma}\left(e_i^\gamma - e_j^\gamma\right) - B_{ij}^{\phi\gamma}\left(f_i^\gamma - f_j^\gamma\right)]
\end{aligned} \quad (1)
$$

where $\Phi = \{a,b,c\}$ denotes the phase set and $\gamma,\phi \in \Phi$; $N = \{0,1,2,...,n\}$ is the bus set, $i,j \in N$; $P_i^\phi$ and $Q_i^\phi$ represent the $\phi$-phase active and reactive power injections at bus $i$; $e_i^\phi$ and $f_i^\phi$ are the real and imaginary parts of the $\phi$-phase voltage at bus $i$; $P_{ij}^\phi$ and $Q_{ij}^\phi$ are the $\phi$-phase active and reactive line flow at branch *i-j*; $G_{ij}^{\phi\gamma}$ and $B_{ij}^{\phi\gamma}$ denote the conductance and susceptance between $\phi$ and $\gamma$ phases at branch *i-j*. Note that if $\phi = \gamma$, then $G_{ij}^{\phi\gamma} = G_{ij}$, $B_{ij}^{\phi\gamma} = B_{ij}$, they are the $\phi$-phase self-conductance and self-susceptance at branch *i-j*. Generally, for balanced power systems, the coupling terms in (1) are ignored and the equations of ACPF for each phase can be simplified without the superscripts.

In the formulations above, we treat $e_i^\phi, f_i^\phi$ as independent variables and $P_i^\phi, Q_i^\phi, P_{ij}^\phi, Q_{ij}^\phi$ as dependent variables with respect to data mining. To facilitate the analysis of power flow equations, we transform (1) into matrix forms:

$$
\begin{aligned}
p_i &= X^T A_i X \\
q_i &= X^T B_i X \\
p_{ij} &= X_{ij}^T A_{ij} X_{ij} \\
q_{ij} &= X_{ij}^T B_{ij} X_{ij}
\end{aligned} \quad (2)
$$

where $p_i, q_i, p_{ij}, q_{ij}$ correspond to $P_i^\phi, Q_i^\phi, P_{ij}^\phi, Q_{ij}^\phi$; $A_i$, $B_i$, $A_{ij}$, $B_{ij}$ are symmetrical but indefinite matrices constituted by entries of the admittance matrix, indicating that all dependent variables in power flow are nonconvex functions of the independent variables. For the balanced three-phase power systems, $X$ denotes the voltage component vector and consists of the single-phase voltage components at all buses, $X = [e_1 \, f_1,..., e_n \, f_n]^T = [x_1 \, x_2,..., x_{2n}]^T$, and $X_{ij}$ denotes another voltage component vector and consists of the single-phase voltage components at buses $i$ and $j$, $X_{ij} = [e_i \, f_i \, e_j \, f_j]^T = [x_{2i-1} \, x_{2i} \, x_{2j-1} \, x_{2j}]^T$. For the unbalanced three-phase power systems, $X$ contains the three-phase voltage components at all buses, $X = [x_1^a \, x_2^a,..., x_{2n}^a, x_1^b \, x_2^b,..., x_{2n}^b, x_1^c \, x_2^c,..., x_{2n}^c]^T$; $X_{ij}$ contains the three-phase voltage components at buses $i$ and $j$, $X_{ij} = [x_{2i-1}^a \, x_{2i}^a \, x_{2j-1}^a \, x_{2j}^a, x_{2i-1}^b \, x_{2i}^b \, x_{2j-1}^b \, x_{2i}^b \, x_{2j-1}^b \, x_{2j}^b, x_{2i-1}^c \, x_{2i}^c \, x_{2j-1}^c \, x_{2j}^c]^T$.

### B. Existing Problems and the Proposed Solutions

The nonlinearity and nonconvexity of the three-phase ACPF (1) introduce challenges to compute optimization and control problems in power systems. Many linearization methods have been proposed to simplify the power flow models and reduce the computation burden, however, at the cost of model accuracy. To overcome these problems, given that the power flow equations are quadratic, data mining techniques can be applied to fit a quadratic convex relationship between the independent and dependent variables, provided that there are large historical databases of power system operations or enough accessible measurements. Theoretically, a convex quadratic approximation of power flows, which is nonlinear, outperforms the linear approximations in terms of model fidelity. Moreover, according to the theory of numerical optimization, a convex quadratic programming is much more computationally tractable than SDP problems.

An important step of obtaining such a DDCQA is to fit a positive semi-definite approximation of the original indefinite coefficient matrix of ACPF. Similar research can be found in other areas like analyzing the correlation matrices of financial stocks employed projection-based algorithms [38]-[40], which converted the original non-convergence problem into a convex optimization problem. Instead, in our study we take full advantage of ensemble learning techniques to approximate to the closest convex representations of power flow. The fitted

3DDCQA of ACPF can be further applied to OPF computation, state estimate or other cases in place of the original nonconvex model.

## III. Ensemble Learning Based Convex Approximation

This section introduces the procedure for the proposed ensemble learning based DDCQA of ACPF in details. First, the nonconvex quadratic mapping in (1) is replaced by a convex quadratic mapping. Then, ensemble learning methods are introduced to infer the parameters of the proposed convex mapping. Finally, the convex mapping is further relaxed into a set of convex constraints with application in OPF.

### A. Convex Mapping Representations

First, we define a convex quadratic mapping in (3) between power, i.e., $p_i$, $q_i$, $p_{ij}$, and $q_{ij}$, and voltage $X$, as the original power flow is represented by a quadratic formulation:

$$\begin{aligned} p_i &= X^T A_i^p X + B_i^p X + c_i^p \\ q_i &= X^T A_i^q X + B_i^q X + c_i^q \\ p_{ij} &= X_{ij}^T A_{ij}^p X_{ij} + B_{ij}^p X_{ij} + c_{ij}^p \\ q_{ij} &= X_{ij}^T A_{ij}^q X_{ij} + B_{ij}^q X_{ij} + c_{ij}^q \end{aligned} \quad (3)$$

where the positive semi-definite coefficient matrices of the quadratic terms at bus $i$ and branch $i$-$j$ are represented by $A_i^*$ and $A_{ij}^*$, respectively; the coefficient vectors of the linear terms are represented by $B_i^*$ and $B_{ij}^*$, and $c_i^*$ and $c_{ij}^*$ are constant terms. Note that here the upper index (*) represents the index set $\{p,q\}$, which corresponds to the active or reactive power. Equations (3) are convex functions, since their coefficient matrices of the quadratic terms are at least positive semi-definite.

### B. Ensemble Learning for Inferring Convex Mapping

Next, ensemble learning, a machine learning technique, is introduced to fit parameter matrices $A_i^*$ and $A_{ij}^*$, vectors $B_i^*$ and $B_{ij}^*$, and constants $c_i^*$ and $c_{ij}^*$ from historical system operation data. Gradient boost (GB) and bagging [37], two typical ensemble learning algorithms, are of particular interest due to their strong ability to enhance basic learners in order to improve the model's performance.

Assume that we are given a training set including $M$ samples $\{(X_m, Y_m)\}_{m=1}^M$. For each sample in an $n$-bus system, the vector of real and imaginary parts vector of bus voltage is denoted by $X_m = [x_{m1}, x_{m2}, ..., x_{m(2n)}]$ for balanced three-phase systems, or $X_m = [x_{m1}, x_{m2}, ..., x_{m(6n)}]$ for unbalanced three-phase systems. If the dependent variable $Y$ is the active or reactive power at each bus $p_i$ or $q_i$, the observation value can be depicted by $Y_m = p_{mi}$ or $q_{mi}$. Similarly, for $p_{ij}$ or $q_{ij}$, we set $Y_m = p_{mij}$ or $q_{mij}$. The following illustrations of GB and bagging are all based on the dependent variable $p_i$ as a general example. Other dependent variables follow the same procedure as $p_i$.

#### 1) Gradient Boosting

Gradient boosting is widely used to develop a strong learner by combining many weak learners in an iterative fashion [20]-[21] for regression and classification problems. It is considered a gradient descent algorithm that can restrain the overfitting effect of the regularization parameters, such as number of iterations and learning rate. The essence of gradient descent is to adjust parameters iteratively to minimize a loss function. It measures the local gradient of the loss function for a given number of iterations and takes steps in the direction of the descending gradient. Once the gradient is zero, we have reached the minimum. The specific loss function for $p_i$ is computed by the mean squared error function as

$$L(p_{mi}, p_{mi}(X)) = \tfrac{1}{2}(p_{mi} - p_{mi}(X))^2 \quad (4)$$

where $p_{mi}$ and $p_{mi}(X)$ are the observed and estimated values of $p_i$, respectively. The detailed procedure of the algorithm is shown below.

---

**Algorithm: Gradient Boosting**

1. Initialize the model with a constant value $p_{mi}^0(X)$

$$p_{mi}^0(X) = arg\min_{\gamma} \sum_{m=1}^M L(p_{mi}, \gamma) \quad (5)$$

where $\gamma$ is the initial constant vector.

2. For $t = 1$ to $T$ where $T$ is the number of learners.

  a. Compute the negative gradient $r_t$ by

$$r_t = -\left[\frac{\partial L(p_{mi}, p_{mi}(X))}{\partial p_{mi}(X)}\right]_{p_{mi}(X) = p_{mi}^{t-1}(X)} \quad (6)$$

  b. Fit a base learner $h_t(X;\theta)$ by

$$\theta_t = arg\min_{\theta} \sum_{m=1}^M L(r_t, h_t(X_m;\theta)) \quad (7)$$

  where $\theta_t$ presents the coefficient vector of $h_t(X;\theta)$ by fitting $r_t$. Here the polynomial regression is adopted to be a basic learner and fit the parameter matrices $A_i^*$ and $A_{ij}^*$, vectors $B_i^*$ and $B_{ij}^*$, and constants $c_i^*$ and $c_{ij}^*$ in (3).

  c. Compute the learning rate $\beta$ by

$$\beta_t = arg\min_{\beta} \sum_{m=1}^M L(p_{mi}, p_{mi}^{t-1}(X_m) + \beta h_t(X_m;\theta)) \quad (8)$$

  Setting a constant learning rate is also allowed. In practice, there is a common pattern that the smaller $\beta$ is, the lower the descent increment is, and the better generalization is achieved. However, the cost of improving the generalization is the reduction of convergence speed.

  d. Update the model.

$$p_{mi}^t(X) = p_{mi}^{t-1}(X) + \beta_t h_t(X) \quad (9)$$

3. Output $p_{mi}^T(X)$

---

#### 2) Bagging

Bagging, called bootstrap aggregating in some references [22]-[23], is designed to improve model stability and accuracy and is applied in classification and regression analysis. As an ensemble technique, it contributes to reducing variance and avoiding overfitting through adjusting the number of bootstraps — a special case of the model averaging approach. The main work of bagging is to draw random samples with replacement and combine a basic learning method to train models. The algorithm is shown below.

**Algorithm: Bagging**

1. For $bt = 1$ to $BT$ where $BT$ is the number of bootstraps.
   a. At the $bt$-th bootstrap, draw $M'$ ($M' \leq M$) random samples with replacement.
   b. Fit a base learner $p_{mi}^{bt}(X;\theta)$ by
   $$\theta_{bt} = arg\min_{\theta}\sum_{m=1}^{M} L(p_{mi}^{bt}, p_{mi}^{bt}(X_m;\theta)) \quad (10)$$
   where $p_{mi}^{bt}$ is the observed value of $p_i$ at the $bt$-th bootstrap and $\theta_{bt}$ represents the coefficient vector of $p_{mi}^{bt}(X;\theta)$ by fitting $p_{mi}^{bt}$. In a similar way, the polynomial regression as the basic learner is introduced to estimate all parameters in equations 3.

2. Output $p_{mi}^{bag}(X)$ by averaging all bootstrap outcomes in
$$p_{mi}^{bag}(X) = \frac{1}{BT}\sum_{bt=1}^{BT} p_{mi}^{bt}(X) \quad (11)$$
where $p_{mi}^{bag}(X)$ is the predictive value of $p_i$ through bagging.

### C. Multicollinearity and Overfitting

Multicollinearity is a situation in which two or more independent variables are strongly linearly correlated, and it can be ignored if the goal is to reach the accurate prediction of model and the model fitting is good enough [41]. Overfitting refers to a model that works well on the training dataset but poorly on the test dataset. The lack of a training dataset can also result in overfitting.

In the field of machine learning, many remedies have been put forward to overcome these challenges, including removing unwanted independent variables by shrinkage and principal component analysis (PCA) based methods [41], enlarging dataset size, and using ensemble learning [23], [24]. To some extent, removing some variables may help improve the accuracy of prediction in some linear models of power flow. However, it may not only completely change the bus number of the power system but may enlarge the model bias and contribute to poor interpretability and applicability in further optimization and control problems. Therefore, to preserve the original power system and make full use of data-driven techniques, increasing the amount of data to learn is a sound and accessible way to relieve the fitting problem. In addition, ensemble learning is applied to avoid overfitting through regularization and resampling techniques.

### D. Convex Approximation of Optimal Power Flow

After fitting the convex formulations of power flows, the data-driven convex approximation for three-phase optimal power flow consists of the objective function and constraints shown in (12) as below.

$$\text{Minimize } \sum_{i \in G}\sum_{\phi \in \Phi}(c_{i0} + c_{i1}p_i^{g\phi} + c_{i2}p_i^{g\phi 2})$$

$$s.t.\begin{cases} X^T A_i^{p\phi} X + B_i^{p\phi} X + c_i^{p\phi} \leq p_i^{g\phi} - p_i^{L\phi} \\ X^T A_i^{q\phi} X + B_i^{q\phi} X + c_i^{q\phi} \leq q_i^{g\phi} - q_i^{L\phi} \\ x_{2i-1}^{\phi 2} + x_{2i}^{\phi 2} \leq \overline{V}_i^{\phi 2} \\ \underline{p}_i^{g\phi} \leq p_i^{g\phi} \leq \overline{p}_i^{g\phi} \\ \underline{q}_i^{g\phi} \leq q_i^{g\phi} \leq \overline{q}_i^{g\phi} \\ p_{ij}^{\phi 2} + q_{ij}^{\phi 2} \leq \overline{S}_{ij}^{\phi 2} \\ X_{ij}^T A_{ij}^{p\phi} X_{ij} + B_{ij}^{p\phi} X_{ij} + c_{ij}^{p\phi} \leq p_{ij}^{\phi} \\ X_{ij}^T A_{ij}^{q\phi} X_{ij} + B_{ij}^{q\phi} X_{ij} + c_{ij}^{q\phi} \leq q_{ij}^{\phi} \end{cases}, \phi \in \Phi \quad (12)$$

where $G$ is the index set of generators; $c_{i0}$, $c_{i1}$, $c_{i2}$ are cost coefficients of the $i$-th generator, and for distribution networks only $c_{i1}$ is needed which denotes the nodal price at substation $i$; the phase subscript $\phi$ implies the constraints of distinct phases are considered; $p_i^{g\phi}$ and $q_i^{g\phi}$ are the $i$-th generator active and reactive power of the phase $\phi$; $p_i^{L\phi}$ and $q_i^{L\phi}$ are the active and reactive power load of the phase $\phi$ at $i$-th bus; $\underline{p}_i^{g\phi}$, $\overline{p}_i^{g\phi}$, $\underline{q}_i^{g\phi}$ and $\overline{q}_i^{g\phi}$ are the lower and upper limits of the $i$-th generator active and reactive power at the phase $\phi$; $\overline{V}_i^{\phi}$ and $\overline{S}_{ij}^{\phi}$ are the upper limits of the $i$-th bus voltage and the branch $i$-$j$ power flow at the phase $\phi$. For the real and imaginary parts of bus voltage, $x_{2i-1}^{\phi}$ and $x_{2i}^{\phi}$ ($e_i^{\phi}$ and $f_i^{\phi}$), we set the imperative constraints added in (12).

The formulations (12) can also be applied to the OPF in balanced three-phase power systems only if the phase superscripts of constraints above are ignored. And the objective function can also be to minimize the overall power loss in distribution networks.

## IV. SIMULATION ANALYSIS

### A. Case selection and Data Sampling

We expected to use real-world data in this research which was, unfortunately, not available at present. As an alternative, the Monte Carlo method was introduced to simulate and generate random data samples of operation measurements, including bus voltage and bus or branch active and reactive power. Different datasets were randomly sampled from diverse power networks, including IEEE 5-, 9-, 57- and 118-bus transmission systems, and IEEE 34-bus distribution system [36], [44], [45]. For randomly sampling datasets, the active and reactive power loads stochastically change around their preset values within an interval [0.6, 1.1]. Each dataset contains up to 50,000 samples to ensure that there are sufficient samples for training models. Generally, a larger sample set is needed to fit the parameters of a bigger system. It has been observed in [14], [15], and [37], that the required minimum empirical sample size is at least twice or six times the number of buses for balanced or unbalanced power flows.

### B. Predictive Performance Comparison

We randomly chose an equal amount of test and training datasets for each case. For instance, both the test and training sets for case 5 and case 9 contain 100 samples, even for each bootstrap in bagging. Next, we fit the convex functions between the active or reactive power and the bus voltage through polynomial regression (PR), gradient boosting (GB), and bagging. The predictive accuracy is indicated by the average root mean square error (RMSE) of the dependent variable, and the performance demonstration of different methods is characterized by comparing the test RMSEs shown in TABLE I.

TABLE I
RMSES of All Methods on Different Cases

| Case | Method | PR | | GB | | Bagging | |
|---|---|---|---|---|---|---|---|
| RMSE (10 e-05) | | Test | Training | Test | Training | Test | Training |
| Case 5 (size=100, T=250, BT=50) | $P$ | 546.56 | 37.39 | 198.18 | 31.03 | 367.31 | 38.42 |
| | $Q$ | 5477.60 | 215.55 | 1330.46 | 209.35 | 2764.26 | 228.99 |
| | $P_{ij}$ | 503.06 | 47.88 | 50.85 | 31.16 | 286.67 | 49.96 |
| | $Q_{ij}$ | 4453.90 | 251.92 | 199.07 | 88.62 | 2376.14 | 246.03 |
| Case 9 (size=100, T=200, BT=50) | $P$ | 4.75 | 1.82 | 3.12 | 1.73 | 4.05 | 2.01 |
| | $Q$ | 3.79 | 1.42 | 2.07 | 1.08 | 3.25 | 1.58 |
| | $P_{ij}$ | 46.14 | 39.35 | 22.78 | 20.21 | 43.32 | 38.57 |
| | $Q_{ij}$ | 93.52 | 77.96 | 84.89 | 68.61 | 86.13 | 79.30 |
| Case 57 (size=200, T=200, BT=50) | $P$ | 401.14 | 3.07 e-10 | 134.31 | 0.33 | 210.74 | 4.76 e-10 |
| | $Q$ | 562.07 | 5.07 e-10 | 185.44 | 0.08 | 305.51 | 8.58 e-10 |
| | $P_{ij}$ | 11.18 | 7.53 | 8.54 | 5.36 | 9.39 | 7.07 |
| | $Q_{ij}$ | 26.29 | 18.28 | 17.41 | 14.14 | 23.57 | 19.24 |
| Case 118 (size=300, T=140, BT=50) | $P$ | 479.20 | 8.07 e-10 | 321.36 | 0.85 | 423.01 | 4.95 e-10 |
| | $Q$ | 440.60 | 3.73 e-10 | 280.41 | 16.90 | 389.50 | 4.82 e-10 |
| | $P_{ij}$ | 62.2 | 0.51 | 21.12 | 2.09 | 57.6 | 0.53 |
| | $Q_{ij}$ | 22.12 | 1.87 | 14.89 | 1.37 | 20.40 | 1.91 |

Note that here T and BT are the maximum numbers of learners or bootstraps in GB and bagging. The unit of data above is 10e-05. From TABLE I, we have the following observations:
- The ensemble learning algorithms, GB and bagging, consistently worked better than PR on all cases;
- GB outperforms bagging and PR on all cases;
- The training RMSE is smaller than the test RMSE for any dependent variable, even though sometimes they seem to be close, as in case 9 and case 57.

C. *Tuning Learning Parameters*

Tuning engineering is a necessity in machine learning applications, showing that the model performance described by the test and training RMSEs is directly associated with a set of regularization parameters to handle overfitting. The following sections focus on the tuning process of GB and bagging, respectively, for the number of learners T and the number of bootstraps BT.

1. *Tuning the Number of Learners T in GB*

In order to determine the number of learners that effectively improve the model performance, all results of the active power injections $p_i$ for case 5, case 9, case 57 and case 118 as paradigms are plotted in Fig. 1 ~ Fig. 4. Each plot depicts the trend of RMSEs (the logarithms of RMSEs) with the increase of the number of learners. Note that the logarithms of RMSEs are used to enlarge the differences between the training and test RMSEs in order to observe and compare them. From these plots, we can observe that:
- For the four cases, the training and test RMSEs of $p_i$ gradually decrease to be stable with increasing the number of learner T.
- Each case reaches the balance point at different numbers of learners. For case 5, the test and training RMSEs tend to be constant after 150 learners have been incorporated. For case 57 and case 118, their test RMSEs hardly change when T=70 and T=80, respectively, while their training RMSEs after T=100 start to drop slightly, with only 0.0001 or 0.001 per additional learner. Similarly, all RMSEs of case 9 drop only 0.0001 with over 180 learners.
- The test RMSE is always larger than the training RMSE no matter how many learners are used. For case 9, though the training and test RMSEs seem to be very close in the same order of magnitude, the statement above still holds;
- No evident overfitting or underfitting problems are observed on any cases through the tuning process.

2. *Tuning the Number of Bootstraps in Bagging*

Fig. 5 ~ Fig. 8 compare the test or training RMSEs of the active power injection $p_i$ on four cases before and after using bagging. Each plot shows the results of each single bootstrap (blue or red curve) and bagging (orange or gray curve) with the increase of BT. Based on these plots, we can infer that:
- The test or training RMSE of bagging tends to be stable with slight fluctuations after sufficient bootstrap iterations. In case 5, when BT=25, both the test and training RMSEs work well. In case 9, when BT $\geq$ 15, test and training RMSEs stabilize at 4 e-05 and 2 e-05, respectively. Similarly, both case 57 and case 118 have steady test and training RMSEs after BT $\geq$ 15.
- The result of every single bootstrap distributes stochastically around the orange (gray) curve, implying that single learner has its unstable weakness.
- Bagging plays an important role in averaging the variances of all single learners and avoiding overfitting.

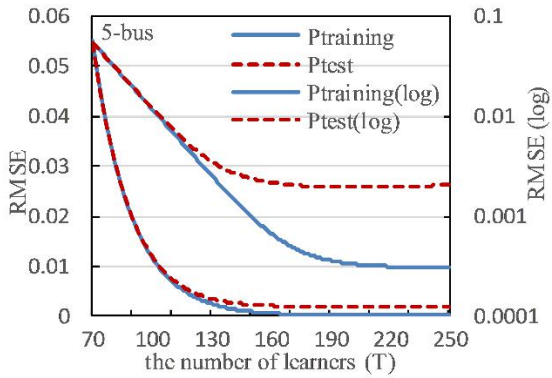

Fig. 1 Case 5: RMSEs of P by Boosting

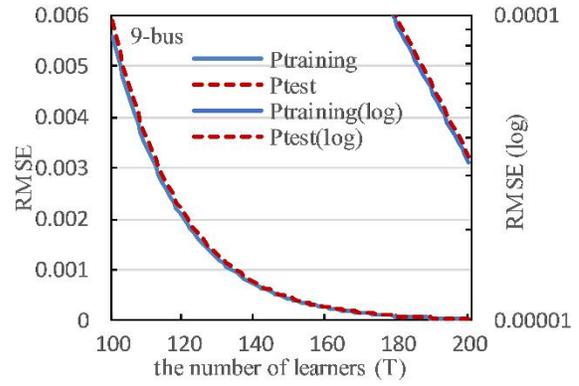

Fig. 2 Case 9: RMSEs of P by Boosting

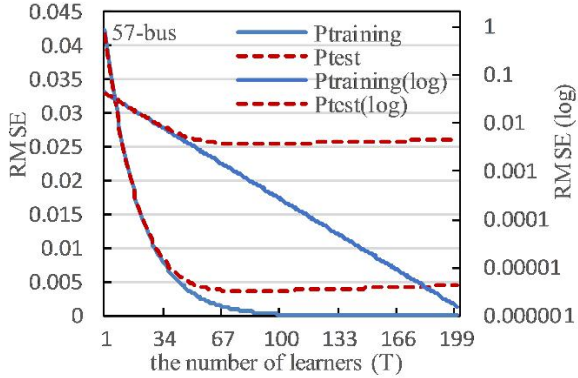

Fig. 3 Case 57: RMSEs of P by Boosting

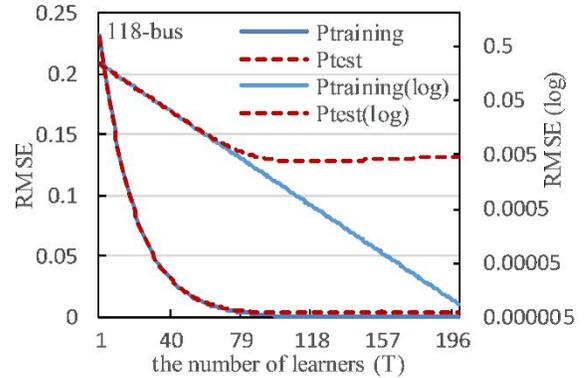

Fig. 4 Case 118: RMSEs of P by Boosting

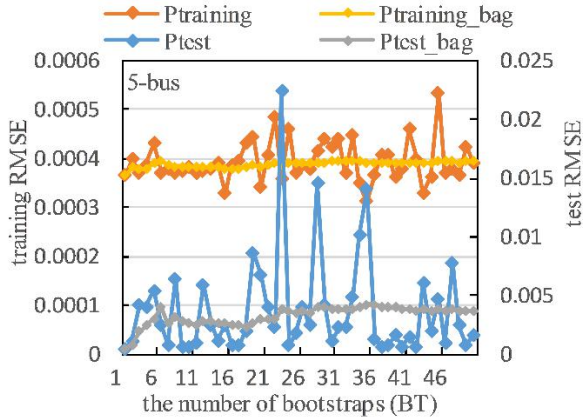

Fig. 5 Case 5: RMSEs of P by Bagging

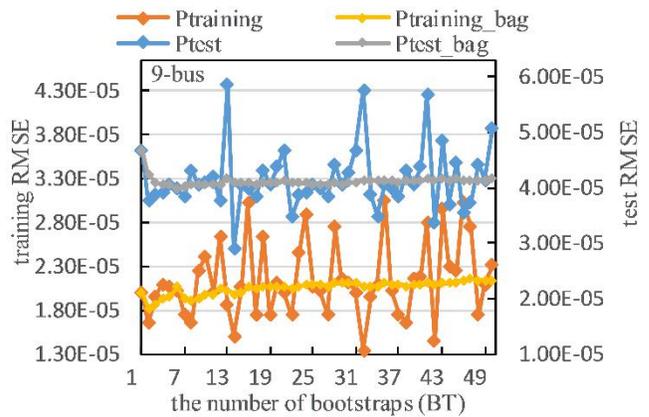

Fig. 6 Case 9: RMSEs of P by Bagging

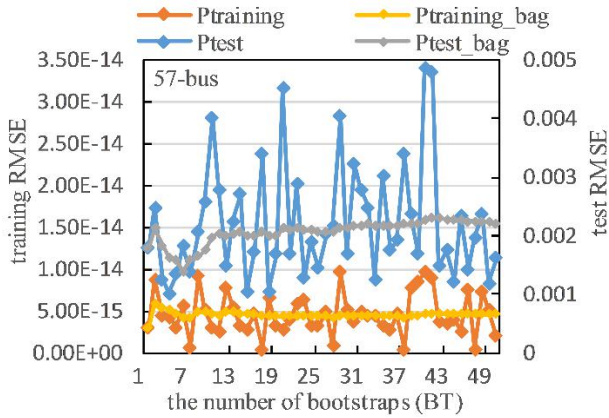

Fig. 7 Case 57: RMSEs of P by Bagging

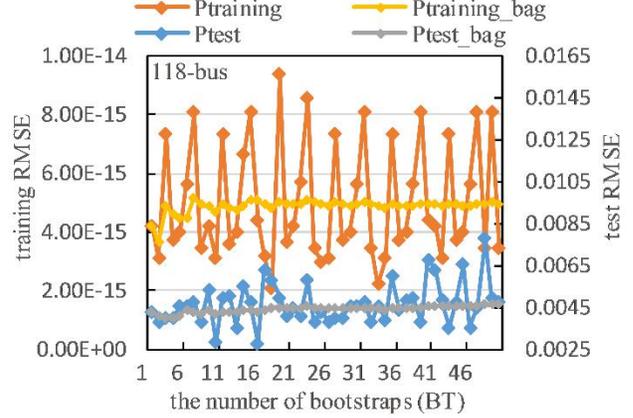

Fig. 8 Case 118: RMSEs of P by Bagging

## D. Computational Evaluation of Algorithms

According to the analysis above, GB exhibits better fitting outcomes than other methods. Based on the convex models fitted by GB, the data-driven convex quadratic approximation of OPF (DDCQAOPF) is applied to compute the minimum generation cost (unit: $/hr) in all cases. In case 34, there are three sets of battery energy storages installed at buses 820, 824 and 860. The battery set at bus 820 is a single-phase source (capacity: 100kVA) and the battery sets at buses 824 and 860 are three-phase sources (single-phase capacity: 200kVA and 100kVA). The power factor of batteries is set at 0.95. The unit electrical energy costs from the substation and batteries are specified at 0.10, 0.12, 0.15, and 0.13 $/kWh. In this case study, we only consider a snapshot as the multi-period load profiles for IEEE 34-bus system are not available. In TABLE II, the results of DDCQAOPF are compared with the original nonconvex ACOPF and the semidefinite programming relaxation of OPF (SDPOPF) [36] to observe the computational accuracy of algorithms. Assume that the results of ACOPF are set as the benchmarks, and the optimality gap, $Err$ shown in TABLE III, is defined as

$$Err = \frac{|OV_{acopf} - OV|}{OV_{acopf}} \times 100\%$$

where $OV_{acopf}$ is the objective value of ACOPF and $OV$ is the objective value of SDPOPF or DDCQAOPF. For comparing the computational efficiency of algorithms, the runtime (unit: second) of each algorithm in different cases is given in TABLE IV. Note that the calculations above are performed through Matlab, cvx package and Mosek solver.

TABLE II~IV reveal that:
- For case 9 and case 118, both SDP relaxation and DDCQA work well, and their objective values are almost the same with ACOPF's.
- Particularly, in some cases (case 5, case 34, case 57) that occur the inexactness of SDP relaxation, DDCQA outperforms SDP relaxation in the computational accuracy.
- A comparison of optimality gaps proves that DDCQA (0%–1.26%) performs more robustly than SDP relaxation (0.01%–13.60%) in the computational accuracy.
- The runtimes indicate that for all cases DDCQA runs more efficiently than SDP relaxation, and its runtimes on case 5 and case 9 are close to ACOPF's.
- DDCQA can be an appropriate alternative of SDP relaxation when it fails to obtain the exact solutions in case 5, case 34 and case 57.

TABLE II
Comparing Objective Values of OPF

| Case | Sample Size | ACOPF | SDPOPF | DDCQAOPF |
|---|---|---|---|---|
| Case 5 | 100 | 17551.89 $/hr | 16635.78 $/hr | 17518.12 $/hr |
| Case 9 | 100 | 5296.69 $/hr | 5297.41 $/hr | 5296.70 $/hr |
| Case 34 | 300 | 211.30 $/hr | - $/hr | 213.96 $/hr |
| Case 57 | 200 | 12100.86 $/hr | 10458.06 $/hr | 12087.96 $/hr |
| Case 118 | 300 | 129660.70 $/hr | 129713.07$/hr | 129454.02 $/hr |

TABLE III
Comparing Optimality Gaps

| Case | Sample Size | ACOPF | SDPOPF | DDCQAOPF |
|---|---|---|---|---|
| Case 5 | 100 | 0% | 5.21% | 0.19% |
| Case 9 | 100 | 0% | 0.01% | 0.00% |
| Case 34 | 300 | 0% | -% | 1.26% |
| Case 57 | 200 | 0% | 13.60% | 0.11% |
| Case 118 | 300 | 0% | 0.04% | 0.16% |

TABLE IV
Comparing Runtimes of Different Methods

| Case | Sample Size | ACOPF | SDPOPF | DDCQAOPF |
|---|---|---|---|---|
| | | Runtime (second) | | |
| Case 5 | 100 | 2.72 | 17.98 | 3.07 |
| Case 9 | 100 | 2.95 | 19.79 | 3.23 |
| Case 34 | 300 | 1.54 | 3.03 | 2.79 |
| Case 57 | 200 | 3.43 | 25.11 | 7.65 |
| Case 118 | 300 | 3.19 | 36.89 | 11.57 |

## V. CONCLUSION AND FUTURE WORK

This paper develops an ensemble learning based DDCQA for power flow. Unlike the most linear approximations of power flow that ignore the interaction terms between the bus voltages, the DDCQA retains the interaction terms and model accuracy. The proposed DDCQA of three-phase ACPF model mainly has two advantages. First, it is more computationally effective than the other available convex model—the SDP relaxation. Second, its accuracy is similar to the SDP relaxation in some cases while outperforms the later one in some other cases. More importantly, the DDCQA, which is learning-based model, has high potential in performance-improvement. On the aspect of machine learning, we introduce an emerging technique—ensemble learning algorithms — to improve the predictive performance of DDCQA. Based on these fitted convex models of power flows, a data-driven convex approximation is proposed and compared with conventional SDP relaxation. Finally, the experimental analysis of IEEE standard systems shows that ensemble learning methods work better than the basic learner, gradient boosting yields the best convex model of power flows, and the proposed convex approximation is superior to SDP relaxation in the computing accuracy and efficiency. Especially, in some cases that the solutions of SDP relaxation are not exact, the proposed algorithm can be an advisable alternative. Our future work will be extended to more practical applications of data-driven convex forms of power flows in control and optimization problems as well as issues with intermittent renewable energy resources.